\documentclass[%
 reprint,
 amsmath,amssymb,
aps,
pra,
]{revtex4-2}

\usepackage{graphicx}
\usepackage{dcolumn}
\usepackage{bm}

\usepackage{caption}
\usepackage{subcaption}

\usepackage{hyperref}
\usepackage{amsmath}
\usepackage{amsthm}
\usepackage{amsfonts}
\usepackage{mathtools}
\usepackage{physics}

\newcommand{\parentheses}[1]{\left(#1\right)}

\newcommand{\Braces}[1]{\left\{#1\right\}}
\newcommand{\angles}[1]{\langle#1\rangle}

\DeclareMathOperator{\diag}{diag}

\renewcommand{\Re}{\operatorname{Re}} %
\renewcommand{\Im}{\operatorname{Im}}

\newcommand{\R}{\mathbb{R}}
\newcommand{\C}{\mathbb{C}}
\newcommand{\N}{\mathbb{N}}
\newcommand{\defeq}{\mathrel{\mathop:}=}

\newcommand{\exval}[1]{{\left\langle#1\right\rangle}}

\newcommand{\Sp}{\mathsf{Sp}}

\newcommand{\LL}{\mathcal L}
\newcommand{\PP}{\mathcal{P}}

\newcommand{\A}{\mathcal A}
\newcommand{\B}{\mathcal B}

\newcommand{\M}{\mathcal M}
\newcommand{\HH}{\mathcal H}

\newcommand{\lrd}{L^2(\R^d)}
\newcommand{\rd}{\R^d}

\begin{document}

\preprint{APS/123-QED}

\title{Semiclassical Wave Packet Dynamics For Nonlinear Schr\"odinger Equations with Harmonic Potentials}

\author{Jared Smith}
\email{Jared.Smith@utdallas.edu}
\author{Tomoki Ohsawa}%
\email{tomoki@utdallas.edu}
\affiliation{%
Department of Mathematical Sciences, The University of Texas at Dallas\\
}%

\date{\today}

\begin{abstract}
We propose an approximate Hamiltonian dynamics for the Gross--Pitaevskii equation $i\hbar\psi_t = -\frac{\hbar^2}{2m}\Delta \psi + V(x)\psi + \hbar^2 \lambda |\psi|^2\psi$ with harmonic potential $V(x)$ and $x\in\mathbb{R}^d$. We do this by extending the Gaussian ansatz of Perez-Garcia et al.~(Phys.~Rev.~A 56, 1424 (1997)) to the semiclassical wave packets of Hagedorn. We derive the evolution equations for the wave packets from the Hamiltonian/symplectic perspective, rather than the Lagrangian/variational one taken by Perez-Garcia et al. We first show that these two perspectives yield identical evolution equations in the Gaussian case. We then derive the evolution equations for an arbitrary element of the semiclassical wave packets. Finally, we demonstrate the efficacy of the approximate dynamics with numerical experiments.
\end{abstract}

\maketitle


\section{Introduction}
Nonlinear Schr\"odinger equations describe a wide range of physical phenomena, including water waves, superconductors, optical fibers, and Bose--Einstein condensates~\cite{karjanto2019nonlinearschrodingerequationmathematical}. In this paper, we will be considering the following Gross--Pitaevskii equation,
\begin{align}
\label{eq:gp}
    i\hbar\psi_t = -\frac{\hbar^2}{2m}\Delta \psi + V(x)\psi + \hbar^2 \lambda \abs{\psi}^2\psi,
\end{align}
where $\lambda \in \R, x \in \R^d$ and $\hbar$ is Planck's constant.

In the linear limit $\lambda \to 0$, the Gross--Pitaevskii equation approaches the linear Schr\"odinger equation. As is well known from the theory of coherent states~\cite{Sc1926,Gl1963,Su1963},  the Gaussian wave packets are exact solutions to the quantum harmonic oscillator. This would suggest that they may be good approximate solutions to~\eqref{eq:gp} with harmonic potential $V(x)$ in the linear limit.

This was notably carried out by Perez-Garcia et al.\cite{Perez-Garcia_1997}, who applied a Gaussian ansatz to~\eqref{eq:gp} in three spatial dimensions with a harmonic trapping potential,
\begin{equation}
\label{eq:pg_potential}
    V(x) = \frac{1}{2}m\nu^2 \sum_{j=1}^d k_j^2 x_j^2,
\end{equation}
where $m,\nu,k_j$ are positive constants.

In this paper, we seek to generalize their approximation of~\eqref{eq:gp} to the non-Gaussian case by means of Hagedorn wave packets~\cite{hagedorn_semiclassical_1980,HAGEDORN199877}. These wave packets are ubiquitous in semiclassical approximations~\cite{lubich2008quantum,Robert2012CoherentSA}, and can be thought of as generalized squeezed states; see~\cite{Ohsawa_2018_hermite,Robert2012CoherentSA} as well as~\cite{VVDodonov_2002,zhang,Perelomov_1977,klauder1985coherent} for additional background.

However, instead of the Lagrangian approach using a variational approximation employed by Perez-Garcia et al., we will be adopting a Hamiltonian perspective, making use of much of the same machinery as in a series of works by Ohsawa~\cite{Ohsawa_2013,Ohsawa_2018_non_gaussian} (see also~\cite{LITTLEJOHN1986193,faou_poisson_2006,pattanayak}) for the linear Schr\"odinger equation. After deriving a set of evolution equations for our approximate dynamics, we shall demonstrate their efficacy with numerical experiments.

\section{Background}
\subsection{Hagedorn Wave Packets and Ladder Operators}
\label{sec:hagedorn}
Here we will give a short overview of the Hagedorn wave packets, so-called for having been first introduced in a series of works by \citet{hagedorn_semiclassical_1980,HAGEDORN199877}. They possess several useful properties which make them desirable for use in semiclassical approximations. In particular, not only do they form an orthonormal basis for the Hilbert space $L^2(\R^d)$ of wave functions on $\R^d$, but also give exact solutions to the (linear) Schr\"odinger equation with arbitrary quadratic potentials; see, e.g., \citet{hagedorn_semiclassical_1980,HAGEDORN199877}, \citet{HellerSemiclassical, HellerSmatrix}, and \citet[Section V]{lubich2008quantum} for more details.

We start by defining the Gaussian wave packet $\varphi_0$ as follows:
\begin{align}
\label{eq:phi0norm}
&\varphi_0(q,p,Q,P;x)\defeq\frac{(\det Q)^{-1/2}}{(\pi\hbar)^{d/4}}\\
&\times \exp\left \{ \frac{i}{\hbar}\left[ \frac{1}{2} (x-q)^TPQ^{-1}(x-q) + p\cdot (x-q) \right ] \right \}, \nonumber 
\end{align}
where $q,p \in \R^d$. The complex $d \times d$ matrices $Q$ and $P$ satisfy the conditions
\begin{equation}
    \label{eq:symplecticity}
    Q^TP - P^TQ = 0, \qquad
    Q^*P - P^*Q = 2i I,
\end{equation}
where $(\,\cdot\,)^*$ stands for the conjugate transpose.

In the present work, where the potential $V(x)$ is assumed to be the harmonic one \eqref{eq:pg_potential}, we may assume that both $Q$ and $P$ are diagonal. This assumption yields simplifications that allow us to obtain closed-form solutions to some of our equations, without compromising the essential properties of the semiclassical wave packets discussed below.
Thus $Q$ and $P$ are essentially vectors in $\C^d$, but we shall continue to write them as complex $d \times d$ matrices, because they must be treated as such for more general potentials and also because the matrix notation is more succinct.

We shall write $y=(q,p,Q,P)$ for the parameter variable as a shorthand, but will occasionally suppress the variable $y$ and write $\varphi(y;x)=\varphi(x)$ for brevity whenever doing so does not cause confusion.

The conditions~\eqref{eq:symplecticity} are commonly referred to as the symplecticity relations on $Q$ and $P$, due to their connection with the symplectic group $\Sp(2d, \R)$; see \cite{Ohsawa_2015_siegel} and references therein for details. They have several important consequences, including
\begin{align*}
\Im(PQ^{-1})=(QQ^*)^{-1}
\end{align*}
and that $(QQ^*)^{-1}$ is symmetric positive definite; note that this must be the case for the Gaussian~\eqref{eq:phi0norm} to be square integrable.

We note in passing that a phase term may also be included in~\eqref{eq:phi0norm} without significantly changing the methodology presented in this work. However, doing so makes calculations more cumbersome. For ease of presentation, we will omit it for now, but we will return to the issue later in the Appendix.

Let $\N_0 \defeq \{0,1,2,\ldots\}$. We will use the following shorthands for multi-index $n = (n_1, \dots, n_d) \in \N_0^d$:
\begin{gather*}
    n!:=n_1!\cdots n_d!,
    \qquad
    \abs{n} := n_1 + \cdots + n_d, \\
    w^n := w_1^{n_1}\cdots w_d^{n_d}
    \quad\text{for}\quad w \in \C^d.
\end{gather*}
For every multi-index $n \in \N_0^d$, we may construct the $n$-th Hagedorn wave packet $\varphi_n$ from the ``ground state'' Gaussian wave packet $\varphi_0$ by means of the ladder operators
\begin{subequations}
\label{eq:ladder_operators}
\begin{align}
     \LL(q,p,Q,P) \defeq-\frac{i}{\sqrt{2\hbar}}[P^T(\hat x - q) - Q^T(\hat p - p)], \label{eq:L}\\
     \LL^\dagger(q,p,Q,P) \defeq \frac{i}{\sqrt{2\hbar}}[P^*(\hat x - q) - Q^*(\hat p - p)],      \label{eq:L*}
\end{align}
\end{subequations}
so that
\begin{equation}
        \begin{split}
        \label{eq:phi_excite}
            \varphi_{n+e_j} \defeq \frac{1}{\sqrt{n_j+1}} \LL_j^\dagger\varphi_n, \\
                \varphi_{n-e_j} \defeq \frac{1}{\sqrt{n_j}} \LL_j\varphi_n,
    \end{split}
\end{equation}
where $e_j$ is $j$th standard basis vector in $\R^d$. Thus, we are able to construct $\varphi_n$ recursively for any multi-index $n$. Note that, given these definitions,
\begin{align*}
    \LL_j\varphi_0 = 0,
\end{align*}
justifying the designation of $\varphi_0$ as the ``ground state'' wave packet.
Analogously, we shall henceforth refer to those $\varphi_n$'s with $|n| \ge 1$ as ``excited states''; note however that those wave packets are usually time-dependent through their parameters and are \textit{not} energy eigenstates of our system of interest.

We similarly define the Hagedorn polynomials $\PP_n(x)$ by
\begin{align}
\label{eq:PP}
    \varphi_n(x) = \frac{\PP_n(x)}{c_n}\varphi_0(x)
    \quad
    \text{with}
    \quad
    c_n \defeq\sqrt{2^\abs{n} n!}.
\end{align}

It is a well-known fact that the Hermite functions constitute another orthonormal basis and the eigenstates for the harmonic oscillator. Their associated polynomials, the Hermite polynomials, are explicitly related to the Hagedorn polynomials by the following formula, due to Ohsawa~\cite{Ohsawa_2018_hermite}:
\begin{align}
\PP_n(x)=\sum_{\substack{k\in\N_0^d \\ \abs{k}=\abs{n}}}\frac{n!}{k!}f^k_n(Q)p_k(\abs{Q}^{-1}(x-q)), \label{eq:hermite}
\end{align}
where $\abs{Q} \defeq (QQ^*)^{1/2}$, $p_k$ is the $k$-th Hermite polynomial, and the coefficients $\{f^k_n(Q)\}_{k,n\in \N_0}$ are defined such that, whenever $\abs{n}=\abs{k}$,
\begin{align}
    (\abs{Q}^{-1}\overline Q w)^k = \sum_{n \in \N_0^d} f^k_n(Q)w^n,\label{eq:fkn}
\end{align}
and $f_n^k(Q) = 0$ for $\abs{n} \neq \abs{k}$.

One may think of the ladder operators~\eqref{eq:ladder_operators} and the Hagedorn wave packets as a generalization of the usual ladder operators and the Hermite functions.
In fact, each Hagedorn wave packet is obtained by applying a certain unitary operator to the corresponding Hermite function; see~\citet{Ohsawa_2018_hermite} for details.

\subsection{Linear Schr\"odinger Equation}
\label{sec:lse}
In this section, we will review the Hamiltonian formalism in the quantum and semiclassical contexts. This formalism is equivalent to the variational approach used in~\cite{Perez-Garcia_1997}, but makes use of the expectation value Hamiltonian $\langle \hat H\rangle$, rather than a Lagrangian. See, for example, \cite{greiner_classical_1996, Talagrand_2026, RevModPhys.38.36ComplexCoordinates,BRODY200119}, and~\cite[Sections 2.2,2.5,2.7, and 3.1-3.3]{JerroldE.Marsden1999}.

Throughout our analysis, we will allow the number of spatial dimensions $d$ to be arbitrary, as doing so will not complicate our analysis. We will also assume a harmonic trapping potential $V(x)$, which allows us to take $Q$ and $P$ to be diagonal. These last two assumptions are essential for deriving the excited state evolution equations in Section~\ref{sec:nlse_excited}.

First, equip $\HH =\lrd$ with the usual right-linear Hermitian inner product 
\begin{align*}
    \ip{\varphi_1}{\varphi_2} = \int_{\rd}\varphi_1^*(x)\varphi_2(x)dx.
\end{align*}
The expectation of the Hamiltonian operator $\hat H$ is given by
\begin{align}
\label{eq:expectation}
\langle \hat H\rangle (\varphi) \defeq \ip{\varphi}{\hat H \varphi},
\end{align}
where
\begin{equation}
\label{eq:Hhat}
\hat H\varphi \defeq -\frac{\hbar^2}{2m}\Delta\varphi + V(x)\varphi.
\end{equation}

Next, we define the Poisson bracket
\begin{align}
\label{eq:poisson}
    \{F,G\} \defeq -\frac{i}{\hbar}\int_{\R^d} \left( \fdv{F}{\varphi}\fdv{G}{\varphi^*} - \fdv{G}{\varphi}\fdv{F}{\varphi^*} \right) dx,
\end{align}
so that setting $\dot G = \Braces{G,\angles{H}}$ for $G=\varphi$ and $G = \varphi^*$ gives us the field Hamilton equations
\begin{subequations}
\label{eq:hamilton_equations}
    \begin{align}
    \label{eq:ham1}
    i\hbar \pdv{\varphi}{t}=\fdv{\angles{\hat H}}{\varphi^*}, \\
    \label{eq:ham2}
    i\hbar \pdv{\varphi^*}{t} = -\fdv{\angles{\hat H}}{\varphi}.
\end{align}
\end{subequations}
From~\eqref{eq:ham1} we get the time-dependent (linear) Schr\"odinger equation
\begin{align}
\label{eq:lse}
    i\hbar \dot \varphi = \hat H\varphi,
\end{align}
and from~\eqref{eq:ham2} we get the complex conjugate of~\eqref{eq:lse}. Note that the bracket~\eqref{eq:poisson} is equivalent to the usual commutator bracket when $F$ and $G$ are both expectation values; see~\cite[Section 3.3]{JerroldE.Marsden1999}.

We now consider the manifold $\M$ defined as the set of parameters $y = (q, p, Q, P)$ for the Gaussian~\eqref{eq:phi0norm}.
In order to adapt the equations~\eqref{eq:hamilton_equations} to the semiclassical setting, we will equip $\M$ with the semiclassical Poisson bracket,
\begin{align}
\label{eq:poisson_ground}
    \{F,G\}_0 &=\pdv{F}{q^j}\pdv{G}{p_j} - \pdv{F}{p_j}\pdv{G}{q^j} \\
    &+ \frac{2}{\hbar}\sum_{k=1}^{2}\parentheses{\pdv{F}{Q^{ij}_k}\pdv{G}{P_{k,ij}} - \pdv{F}{P_{k,ij}}\pdv{G}{Q^{ij}_k}}, \nonumber
\end{align}
where
\begin{equation*}
Q_1 \defeq \Re Q,
\quad
Q_2 \defeq \Im Q,
\quad
P_1 \defeq \Re P,
\quad
P_2 \defeq \Im P.
\end{equation*}
Note that we are adopting the Einstein summation convention of summing over repeated indices. The bracket~\eqref{eq:poisson_ground} is related to the bracket~\eqref{eq:poisson} by
\begin{equation}
\label{eq:poisson_faou}
\begin{split}
        \{F,G\}_0(y) &= \hbar^{-1}DF(y)^T(D\varphi_0(y)^{T}JD\varphi_0(y))^{-1}DG(y)^T \\
        &= \{\hat F,\hat G\}(\varphi_0), 
\end{split}
\end{equation}
where $J \defeq \begin{bmatrix}
    0 && -1 \\ 1 && 0
\end{bmatrix}$ 
is the complex structure (satisfying $J^2 = -I$) on $\lrd \simeq L^2(\R^d,\R) \times L^2(\R^d,\R),$ where $L^2(\R^d,\R)$ is the vector space of real-valued square integrable functions, and $\hat F(\varphi_0) = F(y)$. See~\cite{faou_poisson_2006} for details, as well as~\cite{Ohsawa_2015_siegel} for details on the underlying symplectic geometry. 

We also define the semiclassical Hamiltonian
\begin{align}
    \label{eq:H_0}
    H_0(y) &\defeq \ip{\varphi_0(y;\cdot)}{\hat H\varphi_0(y;\cdot)} \nonumber\\
    &= \frac{p^2}{2m} + V(q) \nonumber\\
    &\quad + \frac{\hbar}{4m}\tr(P P^*) + \frac{\hbar}{4}\tr\left(Q^* D^2V(q) Q\right),
\end{align}
where $D^2V$ stands for the Hessian of potential $V$, but since we are assuming the harmonic potential~\eqref{eq:pg_potential}, the Hessian $D^2V$ is a constant diagonal matrix here.
Then the Hamiltonian system
\begin{equation*}
\dot y_i = \{y_i,H_0\}_0    
\end{equation*}
yields the approximate dynamics
\begin{equation}
\label{eq:lseQP}
    \begin{split}
    \dot q &= \frac{1}{m}p, \\
    \dot p &= -DV(q), \\
    \dot Q &= \frac{1}{m}P, \\
    \dot P &= - D^2V(q)Q,
\end{split}
\end{equation}
where $DV$ stands for the gradient of potential $V$.

The system may be equivalently written as~\footnote{To be precise, the equations~\eqref{eq:lseAB} are obtained from the equations~\eqref{eq:lseQP} by a geometric reduction process; see~\cite{Ohsawa_2015_siegel} for details.}
\begin{equation}
\label{eq:lseAB}
    \begin{split}
    \dot q &= \frac{1}{m}p, \\
    \dot p &= -DV(q), \\
    \dot \A &= -\frac{1}{m}(\A^2-\B^2)-D^2V(q), \\
    \dot \B &= -\frac{1}{m}(\A\B + \B\A),
\end{split}
\end{equation}
where $\A$ and $\B$ are real symmetric matrices defined by $PQ^{-1}=\A + i \B$. We have thus obtained the semiclassical equations of Hagedorn~\cite{hagedorn_semiclassical_1980,HAGEDORN199877} in~\eqref{eq:lseQP}, as well as those of Heller~\cite{HellerSemiclassical,HellerSmatrix} in~\eqref{eq:lseAB}.

A remarkable feature of the Hagedorn wave packets is the following: Denote by $\varphi_n(x,t)$ the $n$-th excited state wave packet whose parameters evolve according to~\eqref{eq:lseQP} or~\eqref{eq:lseAB}. Then $\varphi_n(x,t)$ is an \textit{exact} solution to~\eqref{eq:lse} with harmonic potential $V$.

\section{Nonlinear Schr\"odinger Equation}
\label{sec:nlse}
\subsection{Gaussian Ansatz}
In this section, we will make the same assumptions and effectively repeat the same process as in Section~\ref{sec:lse}, deriving the evolution equations for the Gaussian ansatz~\eqref{eq:phi0norm} from a Hamiltonian perspective, but this time for the Gross--Pitaevskii equation~\eqref{eq:gp} with harmonic trapping potential~\eqref{eq:pg_potential}.

To adapt this method from the linear Schr\"odinger equation to the nonlinear Schr\"odinger equation~\eqref{eq:gp}, we replace the Hermitian operator $\hat H$ with the nonlinear operator $\hat H\varphi + \lambda f(\varphi)$, where $\hat{H}$ is the Schr\"odinger operator~\eqref{eq:Hhat} and $f(\varphi) \defeq \hbar^2 \abs{\varphi}^2\varphi$ is the nonlinear term.
We then define the corresponding energy function analogously to~\eqref{eq:expectation} as follows:
\begin{equation}
    \label{eq:energy}
    \mathcal{E}(\varphi)\defeq\angles{\hat H}(\varphi) + \lambda \int_{\R^d} \frac{\hbar^2}{2}\abs{\varphi}^4dx.
\end{equation}
This is indeed the Hamiltonian function for the nonlinear Schro\"odinger equation~\eqref{eq:gp} in the sense that the Hamiltonian system $\dot{\psi} = \Braces{\psi,\mathcal{E}(\psi)}$ using the Poisson bracket~\eqref{eq:poisson} gives \eqref{eq:gp}; see, e.g.,  \cite{zakharov_complete_1974,mendelson2019rigorousderivationhamiltonianstructure,faddeev_zero_2007}, and~\cite[Section 3]{ashtekar1997geometricalformulationquantummechanics}.

In particular, taking $\varphi$ as the Gaussian $\varphi_0$ from \eqref{eq:phi0norm}, one has the ground state energy
\begin{align}
    \label{eq:E_0}
    E_0(y) &\defeq \mathcal{E}(\varphi_0(y;\cdot)) \nonumber\\
    &= H_0(y) + \lambda\,F_0(y),
\end{align}
where $H_0$ is defined in \eqref{eq:H_0}, and $F_0(y)$ is the contribution of the nonlinear term $f(\varphi_0)$ to $E_0$:
\begin{align}
    F_0(y) &\defeq \int_{\R^d} \frac{\hbar^2 }{2}\abs{\varphi_0}^4 \nonumber \\
    &=\hbar^{2 - d/2}(2\pi)^{-d/2}\abs{\det Q}^{-1}.\label{eq:F0}
\end{align}

Using the Poisson bracket~\eqref{eq:poisson_ground} and the energy function (Hamiltonian) \eqref{eq:E_0}, the Hamiltonian system
\begin{equation*}
    \dot y_i = \Braces{y_i,E_0(y)}_0
\end{equation*}
yields the approximate dynamics 
\begin{equation}
    \label{eq:nlseQPF0}
    \begin{split}
    \dot q &= \frac{1}{m}p, \\
    \dot p &= -DV(q), \\
    \dot Q &= \frac{1}{m}P, \\
    \dot P &= - D^2V(q)Q -\frac{2\lambda}{\hbar} \partial_Q F_0,
\end{split}
\end{equation}
where $\partial_Q$ is the gradient with respect to $Q$.
Substituting~\eqref{eq:F0} into~\eqref{eq:nlseQPF0}, we get
\begin{equation}
\label{eq:nlseQP}
\begin{split}
    \dot q &= \frac{1}{m}p, \\
    \dot p &= -DV(q), \\
    \dot Q &= \frac{1}{m}P, \\
    \dot P &= - D^2V(q) Q, \\ 
    &\quad+ \lambda \hbar^{1-d/2} \frac{\det(QQ^*)^{-1/2}}{(2\pi)^{d/2}} (QQ^*)^{-1}Q^T,
\end{split}
\end{equation}
or, in $(\A,\B)$ variables, 
\begin{equation}
\label{eq:nlseAB}
\begin{split}
 \dot q &= \frac{1}{m}p, \\
    \dot p &= -DV(q), \\
    \dot \A &= -\frac{1}{m}(\A^2 - \B^2) -  \lambda \hbar^{1-d/2} \sqrt{\frac{\det \B}{(2\pi)^d}} \B - D^2V(q), \\
    \dot \B &= -\frac{1}{m}(\A\B + \B\A).
\end{split}
\end{equation}
In either case, the equations of motion are the same as those in Section~\ref{sec:lse}, except for a correction term proportional to $\lambda$ in the $\dot P$ and $\dot \A$ equations. 

We have now obtained the evolution equations in~\cite{Perez-Garcia_1997} as a Hamiltonian system.
Indeed, one obtains the equations~\eqref{eq:nlseAB} from those in~\cite{Perez-Garcia_1997} by the conversion
\begin{equation}
\label{eq:pg_conversion}
\begin{split}
    \A_{\eta\eta} = -2\hbar^2\beta_\eta \\
    \B_{\eta\eta} = -\hbar\omega_\eta^2 \\
    p_\eta = -\hbar^2(\alpha_\eta + 2\beta_\eta \eta_0) \\
    q_\eta = \eta_0,
\end{split}
\end{equation}
where $\eta = x,y,z$.
This shows the equivalence between the variational approach of \cite{Perez-Garcia_1997} and the Hamiltonian one of ours alluded above.

\subsection{Excited States}
\label{sec:nlse_excited}
In this section, we will generalize the construction in the previous section to the case of non-Gaussian wave packets, more specifically the $n$-th Hagedorn wave packets $\varphi_n$ for an arbitrary multi-index $n \in \N_0^d$.
To do this, we shall use the Poisson bracket
\begin{multline}
\label{eq:poisson_excited}
     \{F,G\}_n =\pdv{F}{q^j}\pdv{G}{p_j} - \pdv{F}{p_j}\pdv{G}{q^j}\\ 
     + \frac{2}{\hbar}\sum_{k=1}^{2}(1+2n_j)^{-1}\parentheses{\pdv{F}{Q^{ij}_k}\pdv{G}{P_{k,ij}} - \pdv{F}{P_{k,ij}}\pdv{G}{Q^{ij}_k}}
\end{multline}
for multi-index $n$ corresponding to the $n$-th excited state wave packet $\varphi_n$. As with~\eqref{eq:poisson_ground}, we are using the Einstein summation convention of summing over repeated indices. This was calculated in $(\A, \B)$ parameters by \citet{Ohsawa_2018_non_gaussian}~\footnote{Technically, Ohsawa calculated symplectic forms $\Omega_n$, which we can equivalently write in bracket form by $\Braces{F,H}_n = \Omega_n(X_F,X_H)$.}, but the same methods are readily applied to $(Q,P)$ parameters. In fact, the expression~\eqref{eq:poisson_excited} is simpler than the one calculated by Ohsawa in $(\A, \B)$ parameters. Note also that the bracket~\eqref{eq:poisson_excited} is related to the canonical bracket~\eqref{eq:poisson} by~\eqref{eq:poisson_faou}, but with $\varphi_0$ replaced by $\varphi_n$.

Similarly, we shall define the energy function (Hamiltonian) for the $n$-th excited state as
\begin{align}
    \label{eq:E_n}
    E_n(y) &\defeq \mathcal{E}(\varphi_n(y;\cdot)) \nonumber\\
    &= H_n(y) + \lambda\,F_n(y),
\end{align}
where
\begin{align*}
    H_n(y) &\defeq \ip{\varphi_n(y;\cdot)}{\hat H\varphi_n(y;\cdot)} \\
    &= \frac{p^2}{2m} + V(q) \nonumber\\
    &\quad + \frac{\hbar}{4m}\tr(P \Lambda_n P^*) + \frac{\hbar}{4}\tr\left(Q \Lambda_n Q^* D^2V(q)\right)
\end{align*}
with
\begin{equation*}
    \Lambda_n \defeq \diag(2n_1 + 1, \ldots, 2n_d+1)
\end{equation*}
and
\begin{align*}
    F_n(y) &\defeq \int_{\rd} \frac{\hbar^2}{2}\abs{\varphi_n(x)}^4 dx,
\end{align*}
which we shall evaluate later as it is very tedious.

We are now able to derive the approximate dynamics for an arbitrary excited state. Using the Poisson bracket~\eqref{eq:poisson_excited} and the energy function~\eqref{eq:E_n}, the Hamiltonian system
\begin{equation*}
    \dot y_i = \Braces{y_i,E_n(y)}_n
\end{equation*}
yields
\begin{equation}
\label{eq:nlseQP_exciteF0}
    \begin{split}
    \dot q &= \frac{1}{m}p \\
    \dot p &= -D V(q) \\
    \dot Q &= \frac{1}{m}P \\
    \dot P &= - D ^2V(q) Q -\frac{2\lambda}{\hbar}\Lambda_n^{-1} \partial_Q F_n.
\end{split}
\end{equation}
Note that, apart from the $\partial_Q F_n$ term in~\eqref{eq:nlseQP_exciteF0}, the equations are otherwise the same as~\eqref{eq:lseQP}. This is what we should expect, since the equations~\eqref{eq:nlseQP_exciteF0} approach the equations~\eqref{eq:lseQP} in the linear limit, and those equations govern exact solutions to time-dependent linear Schr\"odinger equations with quadratic potentials; see~\cite[Section V]{lubich2008quantum} and \cite{HAGEDORN199877} for details.

Let us find a closed-form expression for $F_n(y)$.
We start by calculating, using the Hagedorn polynomial $\PP_n$ defined in \eqref{eq:PP},
\begin{align}
    F_n(y)&=\int_{\rd} \frac{\hbar^2}{2}\abs{\varphi_n(x)}^4 dx \nonumber \\
    &= \int_{\rd} \frac{\hbar^2}{2c_n^4}\abs{\PP_n(x)}^4\abs{\varphi_0(x)}^4 dx \nonumber \\
    \label{eq:exciteint1}
    &= \frac{\hbar^2}{2c_n^4} (\pi\hbar)^{-d} \abs{\det Q}^{-2} \int_{\rd}  \abs{\PP_n(x)}^4 \\
    &\times \exp\left\{-\frac{1}{2}\left (\frac{4}{\hbar}(x-q)^T (QQ^*)^{-1}(x-q) \right )\right\} dx. \nonumber
\end{align}
Next, we make the change of variables $\xi = 2(QQ^*)^{-1/2}(x-q)/\sqrt{\hbar}$. Then~\eqref{eq:exciteint1} can be rewritten
\begin{align}
    \label{eq:exciteint2} F_n(y)&=\frac{\hbar^2}{2c_n^4}(\pi\hbar)^{-d} \abs{\det Q}^{-1}\frac{\hbar^{d/2}}{2^d}\\
    &\quad\times  \int_{\rd}  \abs{\hat{\PP}_n(\xi)}^4 
    e^{-\xi^T \xi/2}
    d\xi, \nonumber
\end{align}
where $\hat{\PP}_n(\xi) \defeq \PP_n (\sqrt{\hbar}(QQ^*)^{1/2}\xi/2 + q)$.

Now the integral in~\eqref{eq:exciteint2} is essentially the expectation value of $\abs{\hat{\PP}_n(\xi)}^4$ with respect to the Gaussian $e^{-\xi^T \xi/2}$. Integrals like this come up frequently in the context of Wick's theorem, and have nice formulas in terms of the polynomials being integrated. See~\cite[Section 1.1]{zinn-justin_quantum_2021}, for example. The integral in~\eqref{eq:exciteint2} is thus evaluated, yielding
\begin{align}
    F_n(y)= F_0(y)c_n^{-4}\left[ \abs{\hat{\mathcal  P}_n(\partial_b)}^4 e^{\frac{1}{2}b^Tb} \right]_{b=0}.\label{eq:Fn}
\end{align}
Here we are taking $\abs{\hat{\PP}_n(\partial_b)}^4$ to mean a polynomial of differential operators. Since $b^Tb$ has no cross-terms, we may easily differentiate~\eqref{eq:Fn} by means of, say, Faà di Bruno's formula~\cite{faadibruno} to obtain an explicit formula in terms of $\abs{\hat {\PP}_n(\xi)}^4$:
\begin{align}
\label{eq:fn2}
    F_n(y) = F_0(y) W_n(Q),
\end{align}
where
\begin{equation*}
    W_n(Q)\defeq c_n^{-4}\sum_{k \in \mathbb N_0^d}a^n_k(Q) \frac{k!}{(k/2)!}2^{-\abs{k}/2},
\end{equation*}
and the coefficients $a^n_k(Q)$ are functions of $Q$ defined by $\abs{\hat{\PP}_n(\xi)}^4 = \sum_{k \in \mathbb N_0^d} a^n_k(Q) \xi^k$.
Substituting~\eqref{eq:fn2} into~\eqref{eq:nlseQP_exciteF0}, the evolution equation for $P$ becomes
\begin{equation}
\label{eq:nlse_Pdot_wick_intermediate}
    \dot P = -D^2V(q)Q - \frac{2}{\hbar}(1+2n_j)^{-1}(\lambda W_n \partial_QF_0 + \lambda \partial_Q W_n F_0).
\end{equation}

We can simplify further. It turns out that, when $Q$ and $P$ are diagonal (which is the case in the present work), the $\partial_Q W_n$ term in~\eqref{eq:nlse_Pdot_wick_intermediate} vanishes.
To prove this, we first show that $\PP_n(x)$ factors into a product of one-dimensional Hagedorn polynomials when $Q$ and $P$ are diagonal. That is,
 \begin{equation}
 \label{eq:poly_seperation}
     \PP_n(x_1,\ldots, x_d) = \PP_{n_1}(x_1) \cdots \PP_{n_d}(x_d).
 \end{equation}
 Indeed, if $Q$ and $P$ are diagonal, then the exponential in~\eqref{eq:phi0norm} has no cross-terms. It is then straightforward to show that $\varphi_0(x_1,\ldots,x_d)=\varphi_0(x_1)\ldots\varphi_0(x_d)$. Similarly, since there are no cross-terms in~\eqref{eq:L*}, $\LL^\dagger_j$ acts only on $\varphi_0(x_j)$. In other words,
\begin{align*}
    \LL_j^\dagger \varphi_0(x_1,\ldots,x_d) &= \LL_j^\dagger(\varphi_0(x_1)\cdots\varphi_0(x_d)) \\
    &= \varphi_0(x_1)\cdots \PP_1(x_j)\varphi_0(x_j)\cdots \varphi_0(x_d).
\end{align*}
The result~\eqref{eq:poly_seperation} now follows by induction.
 
As a consequence of~\eqref{eq:poly_seperation},
 \begin{equation}
 \label{eq:poly_Qconst}
     \partial_{Q_{ij}}{\abs{\hat\PP_n(y;\xi)}} = 0.
 \end{equation}
This follows easily from~\eqref{eq:hermite} for $d=1$. Indeed,
\begin{align*}
        \abs{\hat\PP_n(y;\xi)}&= \abs{f^n_n(Q)p_n(\xi)} \\
        &= \abs{f^n_n(Q)}\abs{p_n(\xi)}.
\end{align*}
From~\eqref{eq:fkn},
\begin{align*}
\abs{f^n_n(Q)w^n} = \abs{\abs{Q}^{-1}\overline Q w}^n = \abs{w}^n.
\end{align*}
So $\abs{f_n^n(Q)} = 1$ and $\abs{\hat \PP_n(y;\xi)} = \abs{p_n(\xi)}$, which does not depend on $Q$. The result for arbitrary $d$ then follows from~\eqref{eq:poly_seperation}.

The result~\eqref{eq:poly_Qconst} shows that $a^n_k$ does not depend on $Q$; hence neither does $W_n$, and so the $\partial_Q W_n$ term in~\eqref{eq:nlse_Pdot_wick_intermediate} vanishes. The system~\eqref{eq:nlseQP_exciteF0} now takes its ultimate form
\begin{equation}
\label{eq:nlseQP_excite}
    \begin{split}
    \dot q &= \frac{1}{m}p \\
    \dot p &= -D V(q) \\
    \dot Q &= \frac{1}{m}P \\
    \dot P &= - D ^2V(q) Q \\
    &\quad+ \lambda \hbar^{1-d/2} \frac{\det(QQ^*)^{-1/2}}{(2\pi)^{d/2}} W_n \Lambda_n^{-1} (QQ^*)^{-1}Q^T,
\end{split}
\end{equation}
or in $(\A,\B)$ variables,
\begin{equation}
\label{eq:nlseAB_excite}
\begin{split}
 \dot q &= \frac{1}{m}p \\
    \dot p &= -D V(q) \\
    \dot \A &= -\frac{1}{m}(\A^2 - \B^2) - D^2V(q) \\ 
    &\quad - \lambda \hbar^{1-d/2} \sqrt{\frac{\det \B}{(2\pi)^d}} W_n \Lambda_n^{-1} \B \\
    \dot \B &= -\frac{1}{m}(\A\B + \B\A).
\end{split}
\end{equation}
We can see that these are the same evolution equations \eqref{eq:nlseQP} and \eqref{eq:nlseAB} as for the ground state wave packet, but with a constant multiple $W_n/(1+2n_j)$ in each $j$-th term of the nonlinear contributions.
They indeed recover \eqref{eq:nlseQP} and \eqref{eq:nlseAB} as well because for $n = 0$, one has $\Lambda_0 = I$ and $W_0 = 1$.

\subsection{Stability}
We shall now briefly consider the stability of the evolution equations \eqref{eq:nlseAB_excite}. To mirror the analysis in~\cite[Section 3]{Perez-Garcia_1997}, we will also consider the special case of a spherically symmetric potential, i.e. $k_j=1$ in \eqref{eq:pg_potential}, and $\B=b I$. We will also assume throughout that $n=(\overline n,\ldots,\overline n)$, where $\overline n$ is a positive integer. This is mostly a matter of convenience, as considering arbitrary excited states does not affect the stability analysis in the non-spherical case. It would, however, prohibit us from having a stable equilibrium satisfying $\B = b I$.

Define $\kappa_n \defeq (1+2\overline n)^{-1}W_n\lambda\hbar^{1-d/2}(2\pi)^{-1/2}$. Setting $(\dot \A_0, \dot \B_0) = 0$ in \eqref{eq:nlseAB_excite} yields $\A_0 = 0$ and the following equation for $\B_0,$
\begin{equation}
\label{eq:equilibrium}
    \B_0^2+\kappa_n\B_0\sqrt{\det \B_0} - D^2V(q) = 0.
\end{equation}
If we assume the potential is spherical and that $\B_0 = b_0 I$, then the system~\eqref{eq:equilibrium} reduces to a single equation,
\begin{equation}
\label{eq:equilibrium_spherical}
    b_0^2 + \kappa_n b_0^{(d+2)/2} - m\nu^2 = 0.
\end{equation}

Setting $d=3$, we get
\begin{equation*}
    b_0^2 + \kappa_n b_0^{5/2} - m\nu^2 = 0,
\end{equation*}
which, after setting $n=0$ and applying the conversions~\eqref{eq:pg_conversion}, we arrive at the same equilibrium equation as in~\cite[Section 3]{Perez-Garcia_1997}. Since the $n\neq0$ case only differs from the $n=0$ case by a constant $(1+2\overline n)^{-1}W_n$, we expect the same qualitative behavior regardless of $\overline n$. The same logic would apply to~\eqref{eq:equilibrium} and arbitrary $n \in \N_0^3$.

Setting $d=2$ in~\eqref{eq:equilibrium_spherical} instead yields a quadratic equation with positive root
\begin{equation*}
    b_0=\sqrt\frac{m\nu^2}{1+\kappa_n}.
\end{equation*}
If we linearize~\eqref{eq:nlseAB_excite} at the point $(0,b_0)$, we get
\begin{equation}
\label{eq:eq_spherical_linear}
    \begin{bmatrix}
    \dot a \\ \dot b \end{bmatrix} = \begin{bmatrix}
        0 & 2(1+\kappa_n)b_0 \\
        -2b_0 & 0
    \end{bmatrix}
    \begin{bmatrix}
    a \\  b
    \end{bmatrix}.
\end{equation}
The system~\eqref{eq:eq_spherical_linear} is spectrally stable when $\lambda$ is positive, or when $\lambda$ is negative and satisfies
\begin{equation}
\label{eq:stability_condition}
    \abs{\lambda} < \frac{\sqrt{2\pi}(1+2\overline n)}{W_n}.
\end{equation}

\section{Numerical Results}
\label{sec:num}
We shall assume units such that $\hbar=m=1$ for all of the numerical experiments to follow.
We also set $\lambda = -1$, for which the stability condition~\eqref{eq:stability_condition} is satisfied in all the cases.
The initial wave function is the wave packet $\varphi_n(q_0,p_0,Q_0,P_0)$ defined by~\eqref{eq:phi_excite}, where $(q_0, p_0) = (-1,0)$, and $Q_0$ and $P_0$ are chosen so that $\Im(P_0 Q_0^{-1})$ is close to the equilibrium state $\B_0$ obtained by solving \eqref{eq:equilibrium} numerically. 

\begin{figure*}[ht]
\centering
\begin{subfigure}[b]{0.45\textwidth}
\includegraphics[scale=0.875]{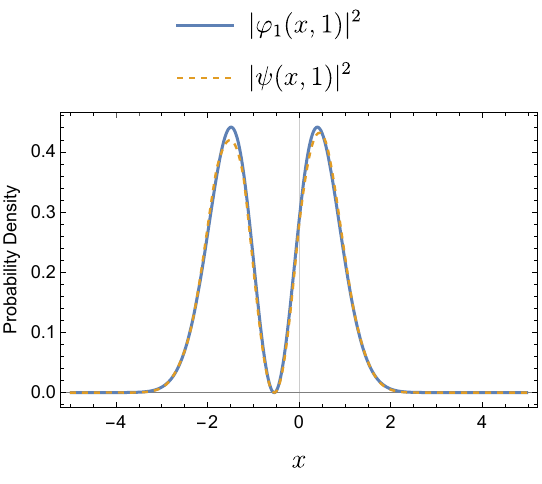}
    \caption{$n=1$}
    \label{fig:prob_density1}
\end{subfigure}
\hfill
\begin{subfigure}[b]{0.45\textwidth}
\includegraphics[scale=0.875]{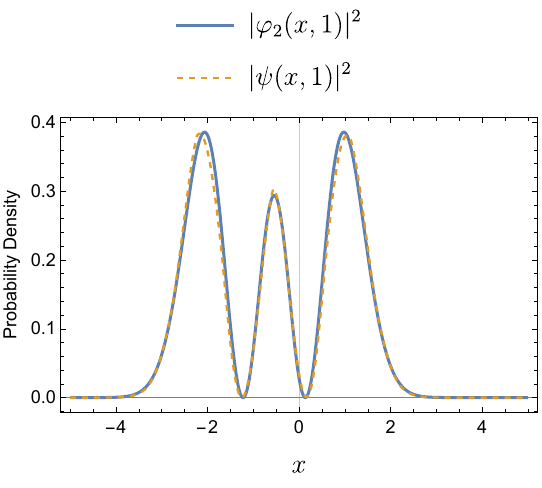}
    \caption{$n=2$}
    \label{fig:prob_density2}
\end{subfigure}
\vspace{3ex}\\
\begin{subfigure}[b]{0.45\textwidth}
\includegraphics[scale=0.875]{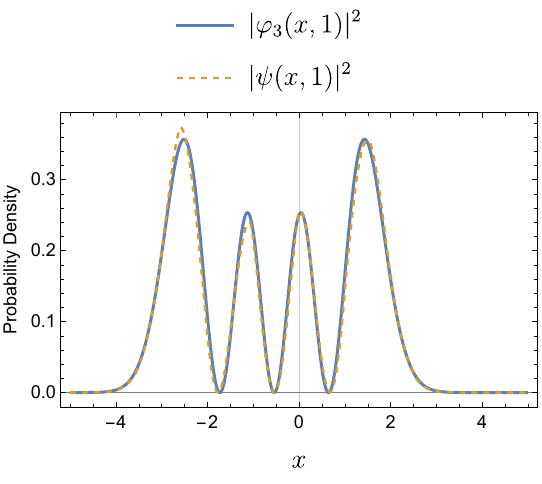}
    \caption{$n=3$}
    \label{fig:prob_density3}
\end{subfigure}
\hfill
\begin{subfigure}[b]{0.45\textwidth}
\includegraphics[scale=0.875]{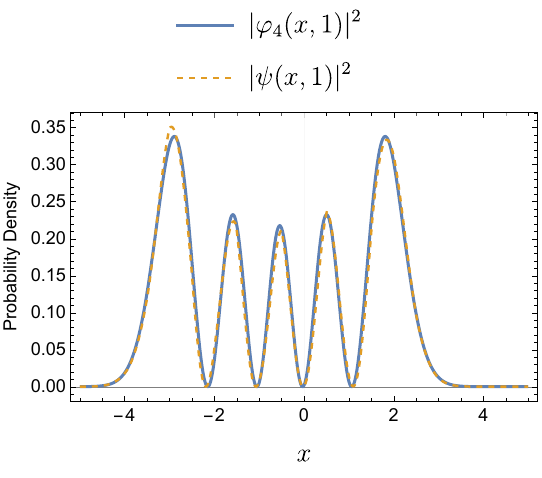}
    \caption{$n=4$}
    \label{fig:prob_density4}
\end{subfigure}
\caption{Probability density of approximate dynamics~\eqref{eq:nlseQP_excite} in one dimension ($d = 1$) (blue) versus numerical solution of~the GP equation~\eqref{eq:gp} (dashed orange) for $\lambda = -1$ at $t=1$.}
\label{fig:density_1d}
\end{figure*}

Let us write the approximate solution as $\varphi_n(x,t) \defeq \varphi_n(y(t);x)$, where $y(t)$ is given by the dynamics~\eqref{eq:nlseQP_excite}.
First, we compare the probability densities of the approximate solution $\varphi_n(x,t)$ with that of a numerical solution $\psi(x,t)$ to the GP equation~\eqref{eq:gp} obtained by a pseudo-spectral method.
In Figure~\ref{fig:density_1d}, we are plotting $|\varphi_n(x,t)|^2$ and $|\psi(x,t)|^2$ at $t=1$ in one spatial dimension ($d = 1$) for several choices of $n$.

\begin{figure}
    \centering
    \includegraphics[width=.95\linewidth]{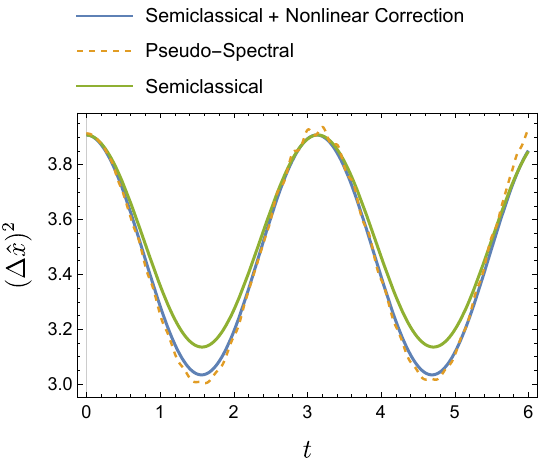}
    \caption{Variance of the approximate dynamics~\eqref{eq:nlseQP_excite} (blue) versus numerical solution of the GP equation~\eqref{eq:gp} (dashed orange) versus the approximate dynamics without the nonlinear correction term (i.e. the semiclassical evolution equations of Hagedorn~\eqref{eq:lseQP}) (green) in one spatial dimension ($d = 1$) for $n=3$ and $\lambda = -1$.}
    \label{fig:1Dvariancehagedorn3n}
\end{figure}

Next, we compare the variances $(\Delta\hat x)^2 = \exval{\hat{x}^2} - \exval{\hat{x}}^2$ of position operator $\hat{x}$ for $\varphi_n(x,t)$ and $\psi(x,t)$. Note that we have a closed-form expression for the variance of the wave packet $\varphi_n(x,t)$, which is especially simple when $Q$ and $P$ are diagonal,
\begin{equation*}
    (\Delta\hat x)^2=\label{eq:variance}
    \frac{\hbar}{2}\sum_{j=1}^d (2n_j +1)(Q\overline Q)_{jj}.
\end{equation*}
In Figure~\ref{fig:1Dvariancehagedorn3n}, we are comparing them for $d = 1$ computed in three different ways: (i)~the wave packet $\varphi_n(x,t)$ whose parameters evolve according to our semiclassical equations~\eqref{eq:nlseQP_excite}, (ii)~the numerical solution $\psi(x,t)$, and (iii)~the wave packet $\varphi_n(x,t)$ using the semiclassical equations of Hagedorn~\eqref{eq:lseQP} instead of~\eqref{eq:nlseQP_excite}, i.e., without the nonlinear correction terms found in the present work.
Since the equations~\eqref{eq:lseQP} are just the equations~\eqref{eq:nlseQP_excite} without the nonlinear correction term, this comparison demonstrates the significance of the nonlinear correction term derived in this work.

\begin{figure}[ht]
\centering
\begin{subfigure}[b]{0.475\textwidth}
\includegraphics[scale=0.85]{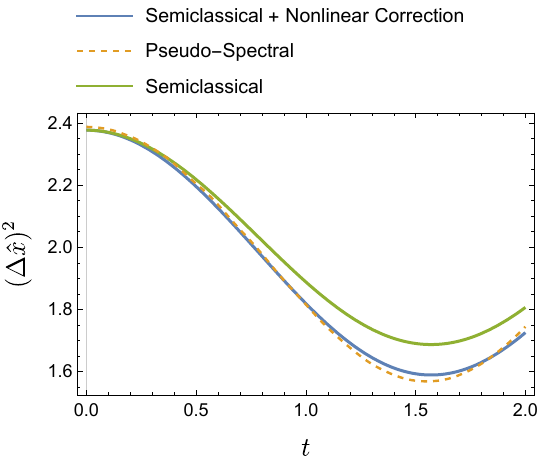}
    \caption{$n=(0,1)$}
    \label{fig:var01}
\end{subfigure}
\vspace{5ex}\\
\begin{subfigure}[b]{0.475\textwidth}
\includegraphics[scale=0.85]{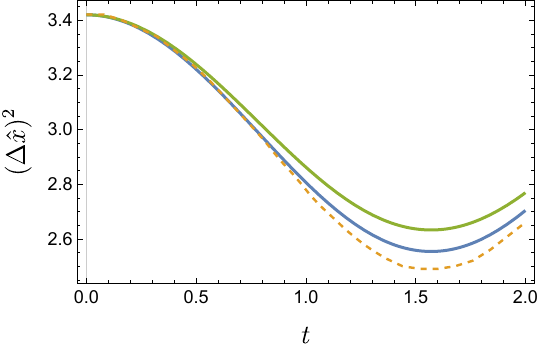}
    \caption{$n=(1,1)$}
    \label{fig:var11}
\end{subfigure}
\caption{Variance of the approximate dynamics~\eqref{eq:nlseQP_excite} (blue) versus numerical solution of the GP equation~\eqref{eq:gp} (dashed orange) versus the approximate dynamics without the nonlinear correction term (i.e. the semiclassical evolution equations of Hagedorn~\eqref{eq:lseQP}) (green) in two spatial dimension ($d = 2$) for $\lambda = -1$.}
\label{fig:variance}
\end{figure}

\begin{figure*}[ht]
\centering
\begin{subfigure}[b]{0.495\textwidth}
\includegraphics[width=\linewidth]{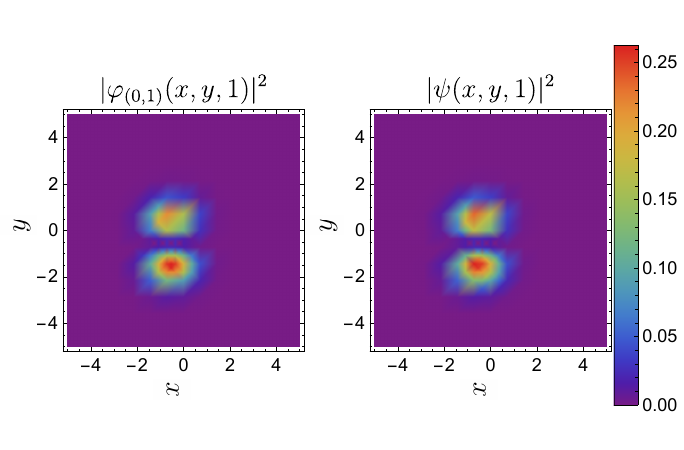}
    \caption{$n=(0,1)$}
    \label{fig:density01}
\end{subfigure}
\begin{subfigure}[b]{0.495\textwidth}
\includegraphics[width=\linewidth]{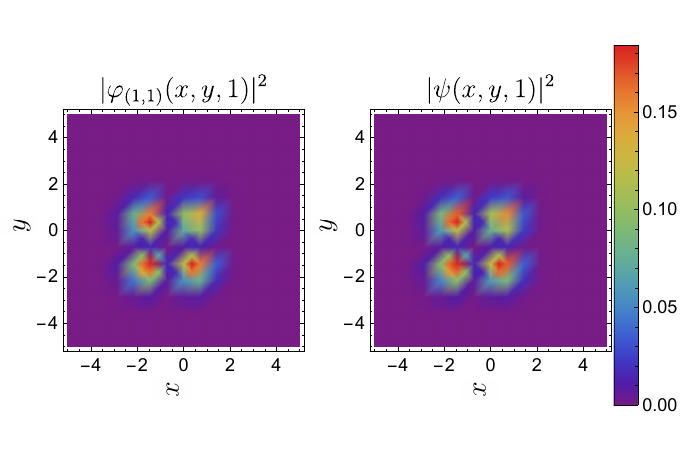}
    \caption{$n=(1,1)$}
    \label{fig:density11}
\end{subfigure}
\begin{subfigure}[b]{0.495\textwidth}
\includegraphics[width=\linewidth]{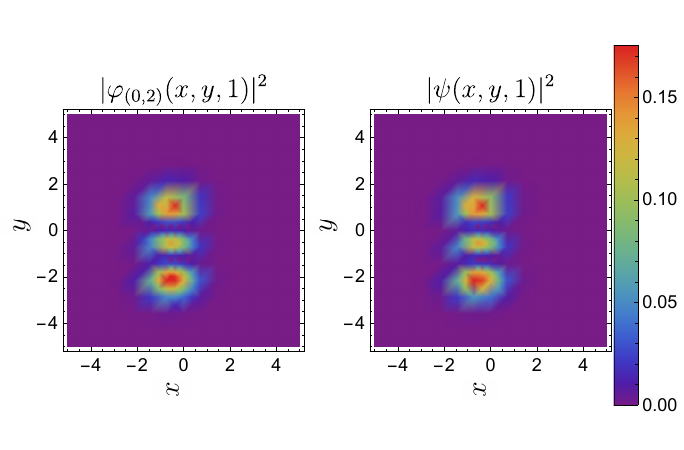}
    \caption{$n=(0,2)$}
    \label{fig:density02}
\end{subfigure}
\begin{subfigure}[b]{0.495\textwidth}
\includegraphics[width=\linewidth]{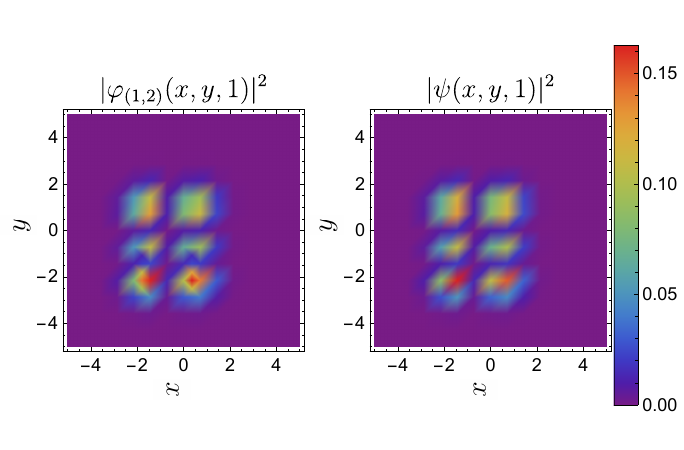}
    \caption{$n=(1,2)$}
    \label{fig:density12}
\end{subfigure}
\begin{subfigure}[b]{0.495\textwidth}
\includegraphics[width=\linewidth]{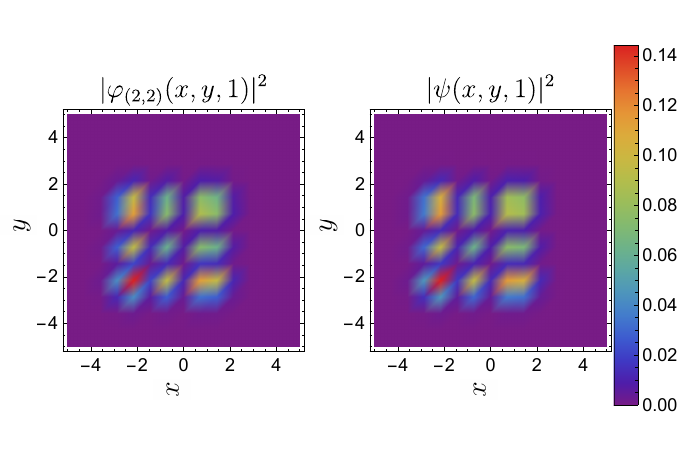}
    \caption{$n=(2,2)$}
    \label{fig:density22}
\end{subfigure}
\caption{Probability density of approximate dynamics~\eqref{eq:nlseQP_excite} in two dimensions (left) versus numerical solution of the GP equation~\eqref{eq:gp} (right) for $d = 2$ and $\lambda = -1$ at $t=1$.}
\label{fig:density_2d}
\end{figure*}

\begin{figure*}[!htbp]
\centering
\begin{subfigure}[b]{0.495\textwidth}
\includegraphics[width=0.85\linewidth]{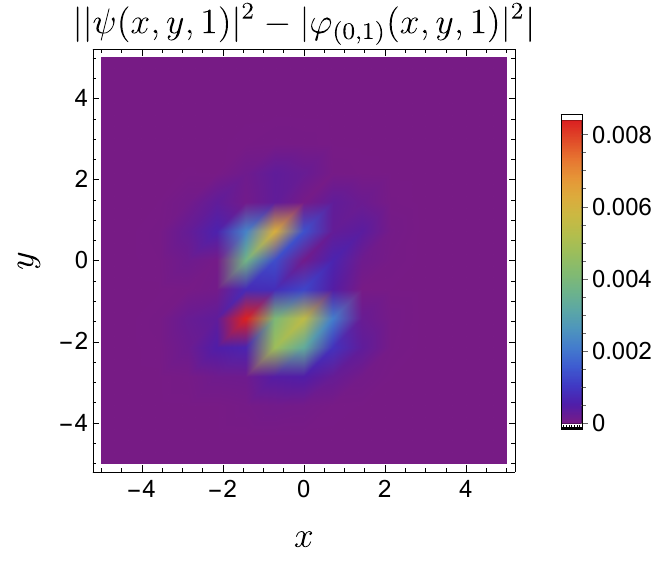}
    \caption{$n=(0,1)$}
\label{fig:density01diff}
\end{subfigure}
\begin{subfigure}[b]{0.495\textwidth}
\includegraphics[width=0.85\linewidth]{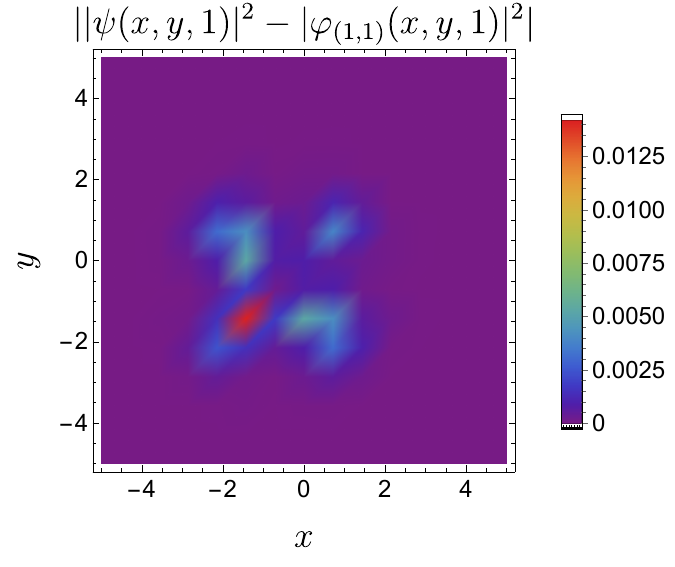}
    \caption{$n=(1,1)$}
    \label{fig:density11diff}
\end{subfigure}
\vspace{0.5ex}\\
\begin{subfigure}[b]{0.495\textwidth}
\includegraphics[width=0.85\linewidth]{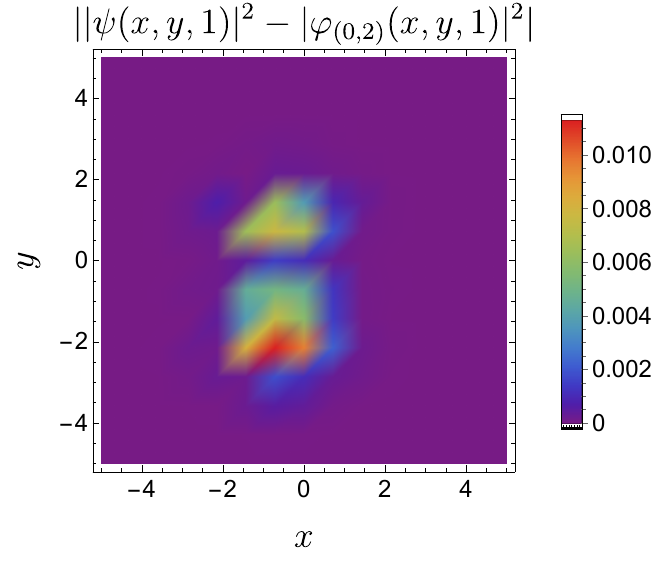}
    \caption{$n=(0,2)$}
    \label{fig:density02diff}
\end{subfigure}
\begin{subfigure}[b]{0.495\textwidth}
\includegraphics[width=0.85\linewidth]{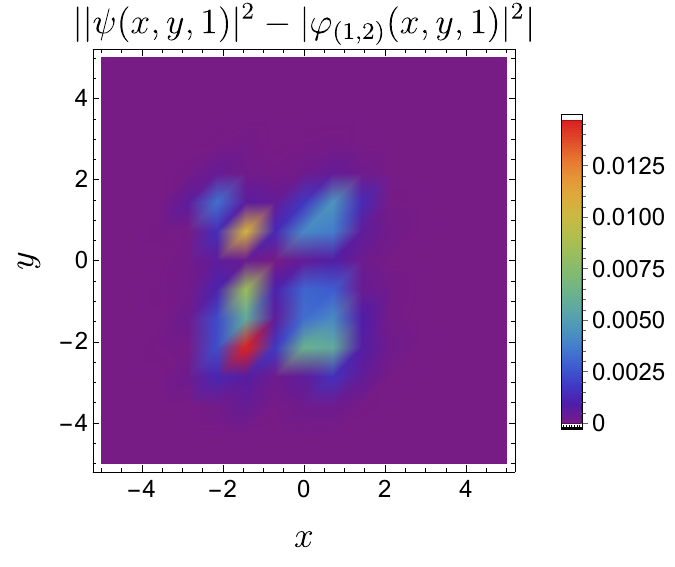}
    \caption{$n=(1,2)$}
    \label{fig:density12diff}
\end{subfigure}
\vspace{0.5ex}\\
\begin{subfigure}[b]{0.495\textwidth}
\includegraphics[width=0.85\linewidth]{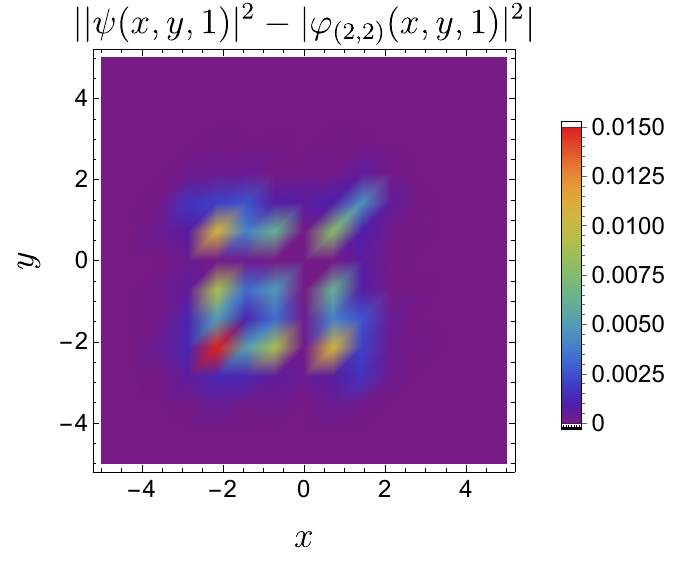}
    \caption{$n=(2,2)$}
    \label{fig:density22diff}
\end{subfigure}
\caption{Absolute difference of probability densities $\abs{\abs{\psi(x,t)}^2-\abs{\varphi_n(x,t)}^2}$ for $d = 2$ and $\lambda = -1$ at $t=1$, where $\varphi_n(x,t)$ is given by the approximate dynamics~\eqref{eq:nlseQP_excite}, and $\psi(x,t)$ is the numerical solution of the GP equation~\eqref{eq:gp}.}
\label{fig:density_2d_diff}
\end{figure*}

Figures~\ref{fig:density_2d} and~\ref{fig:variance} are showing the same thing for two spatial dimensions ($d = 2$) for several choices of $n$. In Figure~\ref{fig:density_2d_diff}, we are also plotting the absolute difference of probability densities $\abs{\abs{\psi(x,t)}^2-\abs{\varphi_n(x,t)}^2}$ at $t=1$.

\section{Conclusion}
In this work, we derived approximate dynamics for the Gross--Pitaevskii equation~\eqref{eq:gp} with harmonic potential. Not only were we able to re-derive the evolution equations for the Gaussian Ansatz in~\cite{Perez-Garcia_1997}, but we were also able to extend the approximation method to the Hagedorn wave packets of arbitrary index in any number of dimensions.

Furthermore, as demonstrated by the numerical experiments, the approximate dynamics show excellent agreement with numerical solutions of the GP equation~\eqref{eq:gp} given by a pseudo-spectral method.

In semiclassical mechanics, the linear Schr\"odinger equation with quadratic potential $V$ is solved exactly by the Hagedorn wave packets whose parameters evolve according to the equations~\eqref{eq:lseQP}. This is generalized to an approximation method for the linear Schr\"odinger equation with a broader class of potentials $V$ by allowing for superpositions of wave packets whose parameters all evolve according to~\eqref{eq:lseQP}; see~\cite{HAGEDORNIII,HagedornIV}, and~\cite[Section V]{lubich2008quantum} for details. Since the equations~\eqref{eq:lseQP} for linear Schr\"odinger equations are generalized to the equations~\eqref{eq:nlseQP_excite} for nonlinear Schr\"odinger equations, it would be natural to ask whether this allows us to similarly generalize to a broader class of potentials $V$ for nonlinear Schr\"odinger equations.
However, for such potentials, we cannot assume that $Q$ and $P$ are diagonal any more; this makes it difficult to obtain some closed-form expressions, particularly the integral in $F_n(y)$ from \eqref{eq:exciteint1}.
We leave this question for future work.

\section{Acknowledgments}
T.O. was supported by NSF grant DMS-2006736.
We would like to thank Jimmie Adriazola for suggesting the paper~\cite{Perez-Garcia_1997} that inspired this research idea, and Roy Goodman for helpful comments and discussions.

\appendix*
\section{Phase Dynamics}
The most straightforward way to incorporate a phase term into the procedure outlined in this work would be to replace the Gaussian~\eqref{eq:phi0norm} with
\begin{equation}
\label{eq:phi0extra}
\begin{split}
    \varphi_0(q,p,Q,P,\phi,\delta;x) =
    \exp\biggl\{&
    \frac{i}{\hbar}\biggl[\frac{1}{2}(x-q)^TPQ^{-1}(x-q) \\
    &+ p\cdot(x-q) + (\phi + i \delta)\biggr] \biggr\},
\end{split}
\end{equation}
which includes a normalization term $\delta$ and a phase $\phi$. Inclusion of the normalization term is necessary in the Hamiltonian formulation, since $\phi$ and $\delta$ are canonically conjugate. Using~\eqref{eq:phi0extra} as the ground state Hagedorn wave packet, we construct arbitrary $\varphi_n$ just as we did in Section~\ref{sec:hagedorn}.
We can follow the same method outlined in Section~\ref{sec:nlse} with the two extra variables.

The main drawback of this approach is that the bracket structure given by~\eqref{eq:poisson_faou} is much more complicated; see \cite{Ohsawa_2013}. However, we are able to write down the evolution equations in terms of the other dynamical variables $(\dot q, \dot p, \dot Q, \dot P)$, similarly to what is done in~\cite{Perez-Garcia_1997}.
We leave the investigation of the effect of the phase dynamics to future work.

See~\cite{Ohsawa_2013,Ohsawa_2018_non_gaussian} for similar calculations to what is described here with the linear Schr\"odinger equation, using symplectic forms instead of Poisson brackets.

\bibliography{reference}

\end{document}